# Does Gaming Help Improve Cognitive Skills?

Mohnish Chakravarti[1], Arati Chakravarti
Pace Junior Science College, Borivali, Mumbai, India

Email: mohnishchakravarti7@gmail.com

**Summary**
A nationally representative study of video game play among adolescents in the United States showed that 97% of adolescents aged 12 to 17 years play computer, web, and portable (or console) video games (Lenhart et al., 2008). We hypothesized that if people play games as a regular exercise regime, gaming will correlate with an improvement in their cognitive skills. For this experiment, a few games that tested the logical reasoning and critical analysis skills under a given time constraint were coded in Python using Pygame and were played by a group of 7th grade students. In order to test whether there is a relationship between gaming and test performance, we divided the students into two groups and gave them tests before and after the experimentation period in order to measure their improvement. One group played the games while the other did not. In the group of students that played the games, an average improvement of 62.19% was seen ($p < 0.0001$). The group that did not play the games only improved their performance by an average of 18.51% ($p = 0.0882$).

**Introduction**
A nationally representative study of video game play among adolescents in the United States showed that 97% of adolescents aged 12 to 17 years play computer, web, and portable or console video games (Lenhart et al., 2008). In terms of frequency, 31% of adolescents play video games every day and another 21% play games 3 to 5 days a week. Similarly, Gentile et al., conducted a large survey study in the United States and found that 88% of youths aged 8 to 18 years play video games and that the average amount of time spent playing video games per week is 13.2 hours. A recent study showed that older people completing visual processing training had more improved cognitive skills than another group of old people playing crosswords for the same time (Wolinsky et al., 2013). We hypothesized that if students played strategic games as a regular brain-training regime, there would be a measurable improvement in their cognitive skills. The improvements associated with gaming would be possibly due to regularly exercising skills in a competitive environment. Gaming exercises could build critical analysis and logical reasoning skills.
 Previous research in this field has focused on the negative outcomes of playing video games, much less on the positive outcomes.
The volunteers for this study were from the seventh grade of a local school. All the experiments took place on the school premises during school hours with permission from the principal and vice-principal of the school. Two tests of logical reasoning were carried out: one prior to the experiment and the second after the end of the experimentation period of one week to check whether there were improvements in the students' performance. In order to test for a relationship between the test performance of the students and the gaming period, one group was made to take the tests but was not allowed to play the games. Another group was made to take the tests and play the games for the duration of the experiment.

The difficulty level of the tests was the same, and both tested the logical reasoning and critical thinking skills of the children. The games were developed in the Python programming language with the help of Pygame. The games were a hybrid between Othello and Go, two popular board games. They tested the students' strategic skills under a given time constraint.

The results suggested that there is a strong relationship between gaming and improvement in tests score. This supports the hypothesis that gaming and improvement in cognitive skills are strongly correlated to each other. Furthermore, a widespread implementation of such an activity is feasible, so that many students can benefit from this activity.

**Results**

The students were made to play the game in the first period of their school for 30 minutes. On the first day of the experiment, the students were first given a test that they would have to solve in 30 minutes. This exam tested the students' logical reasoning and critical analysis skills and was appropriate for 7th grade students. After the end of the experimentation period, (i.e. on the last day) the students were again given a test. The time limit and the difficulty was the same as that of the first exam.

A total of 45 students took both tests. Students that wrote the first test but did not take the second test and vice-versa were omitted from the study. Their scores were not taken into account while calculating the averages.

In order to test whether there is a relationship between the test performance and the gaming period, the students were divided into 2 groups of 25 and 20 students each. The group of 25 students was made to play the games designed by the author, while the group of 20 students was not. This helps eliminate uncertainties in conclusion. Both groups performed on an average level on the first test, with only 5 students scoring over 75%.

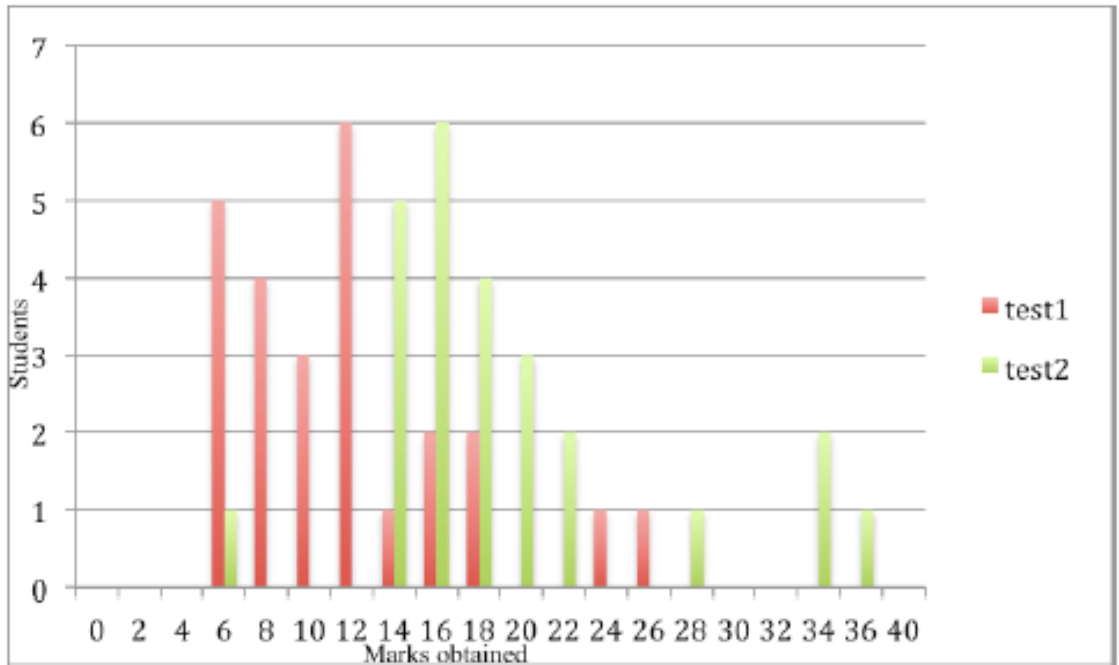

**Figure 1.** This figure represents the data of the gaming group. The Y-axis represents the number of students and the X-axis represents the marks. The graph in red shows the marks obtained in the first test while the graph in green represents the marks obtained in the second test.

Three of those students were from the group that was allowed to play the games and two were from the control group.

Students that played the game
Three students obtained over 75% marks on the second test – something that was not achieved in the first test. The group average for the first test was 11.84, whereas it increased to 19.20. An improvement of 62.19% was seen in the overall performance of the group (Figure 1). The student who scored the highest in the first test did not score the highest on the second, missing out by a single question. Boys out-performed the girls by scoring on average almost two marks higher than the girls. Four people who scored six on the first test improved their scores in the second test.

$$t = \frac{d - d^0}{s_d/\sqrt{n}}$$

**Equation 1.** Calculate the t-value for the paired t-test.

A paired t-test (Equation 1) was used to determine the statistical significance of the data. A paired sample t-test is used to determine whether there is a significant difference between the average values of the same measurement under two different conditions. Both measurements are taken on each unit in a sample, and the test is based on the paired differences between these two values.

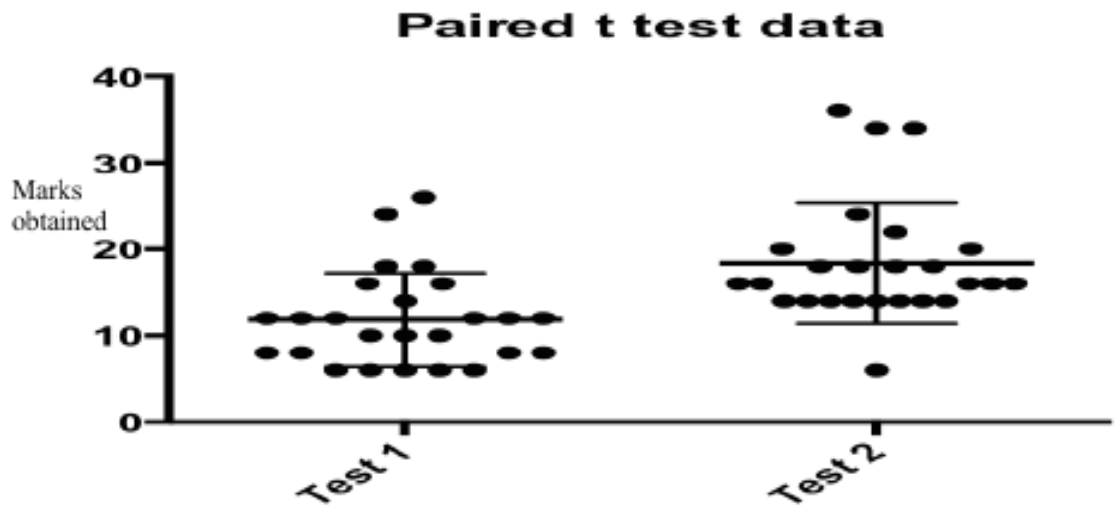

**Figure 2.** This figure represents the mean and the standard deviations of the tests taken by the gaming group.

Let the null hypothesis be that playing games does not improve cognitive skills and let the alternative hypothesis be that it does. The threshold value is set at 0.05. The p-value was found to be less than 0.0001, which is extremely statistically significant. The standard deviation of the differences was found to be 4.744. Figure 2 shows the result of the paired t-test data. The t-value was 6.914 with 24 degrees of freedom.

Students that did not play the game
Only one student achieved more than 75% marks. The group average for the first test was 13.5, whereas it rose to 16 in the second test. An improvement of 18.51% was seen in the performance of the group (Figure 3). The student who scored the highest in the first test also scored the highest in the second test, getting 1 more question right. The performance of the boys and the girls was almost similar, with boys scoring almost 1 mark higher than the girls on average. Both of the students who scored the lowest on the first test scored the lowest on the second test.

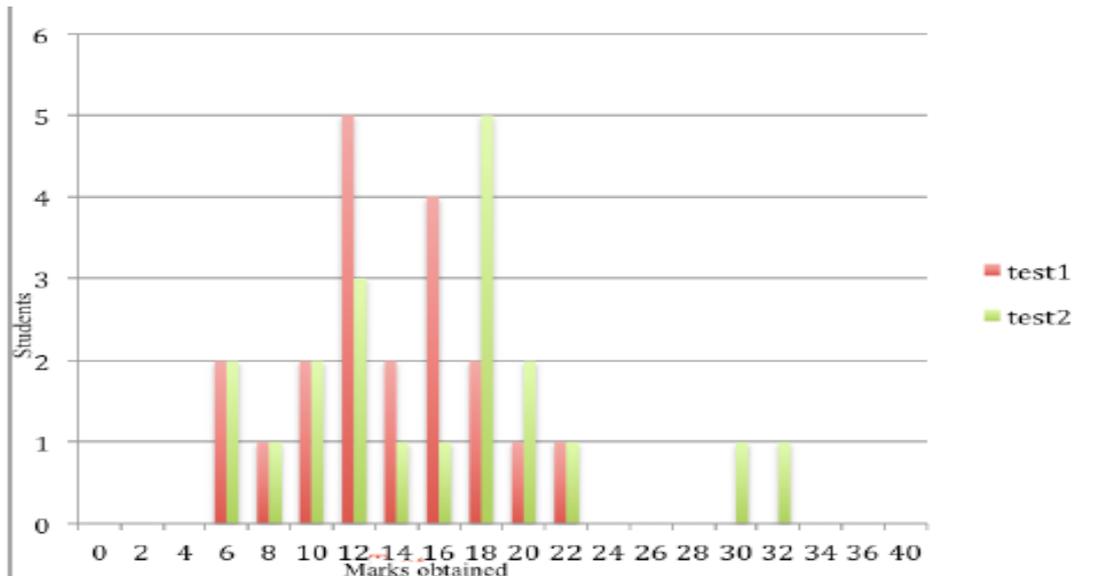

**Figure 3.** This figure represents the data of the non-gaming group. The Y-axis represents the number of students and the X-axis represents the marks. The graph in red shows the marks obtained in the first test while the graph in green represents the marks obtained in the second test.

The paired p-test used to determine the significance of the data is similar to the paired p-test mentioned above. Here, the p-value was found to be 0.0882 which is greater than the threshold value and hence, not statistically significant. The standard deviation of the differences was found to be 6.220. The t-value was 1.798 with 19 degrees of freedom (Figure 4).

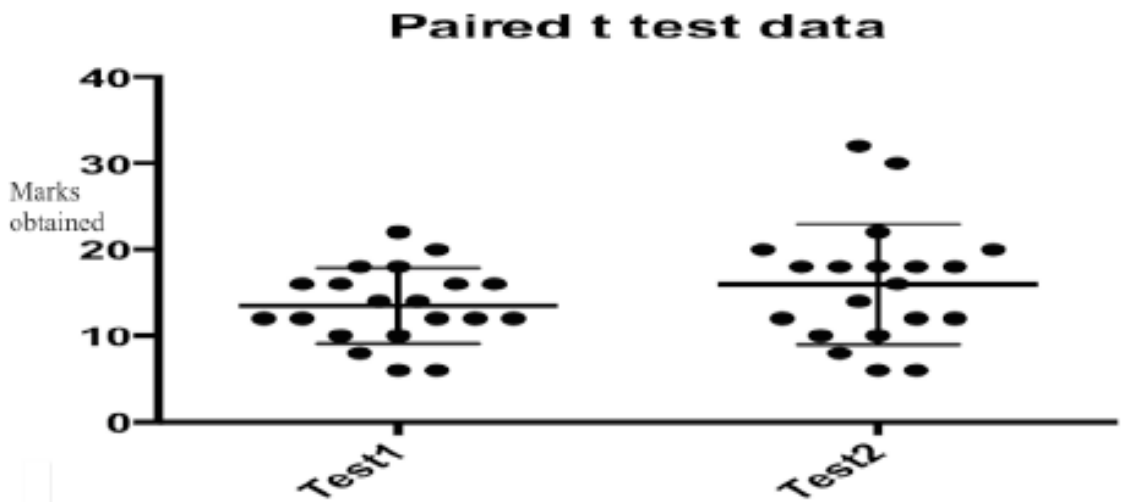

**Figure 4.** This figure represents the mean and the standard deviations of the tests taken by the non-gaming group.

## Discussion

After comparing the data from the group that played the games and the group that did not, a substantial amount of improvement is seen in the performance of the group that played the games.
It was hypothesized that gaming and cognitive skills are strongly correlated with each other. The games tested the logical reasoning and critical analysis skills of the students. Therefore, we propose that daily practice in a competitive environment would eventually result in an improvement of the aforementioned skills.

A total of 45 students took part in the experiment. The students were divided into two groups. The first group was made to play the games and the second group was not made to play the games. In the group that was made to play the games, an improvement of 62.19% was seen on average. In the group that did not play the games, only an improvement of 18.51% was seen on average. Better performance was observed in the second test taken by the gaming group. We hypothesize that this is because the first group had constantly been practicing their logical reasoning and critical analysis skills in a competitive environment by playing the games that were designed as a part of this experiment. Another factor is that, in the games, the computer opponent forced them to critically analyze all the possible outcomes. This might have led to the students being more and more careful while answering the test and also being able to critically analyze all the questions to get the correct answer on the test. The second group was not made to practice daily in a competitive environment, and we propose that it is for this reason that their performance was not as good as that of the gaming group.

It was also seen that boys in general performed better than the girls on both the tests in both the groups. The students in both groups who scored the highest in the first test also saw an increase in their marks in the second tests. Out of the five students in the gaming group who scored the least in the first test, four students improved their performance by at least three marks on average, whereas the remaining student did not see an improvement in her performance. In the group that did not play the games, the students who scored the worst did not see an improvement in their performance.

The students did not report playing strategic video games daily. Only a few students who took part in the experiment played video-games daily, but they only played for leisure purposes. The students also reported that they have not been introduced to a game similar to the one used in the experiments prior to the experimental period. We speculate that had the students played strategic games like the one used in the experiment, it would have resulted in improvement in their test performance, as they would have had more practice in a competitive environment than their peers taking the same test.

The students mentioned that the games did compel them to think before making a move. The game that the students were made to play forced them to make decisions based on their logical reasoning and critical analysis skills under a time constraint – something students also face when they take an exam. As the tiles could be converted back and forth, students had to use their logical reasoning and critical analysis skills to carefully make a move that would increase their chances of preventing the computer from taking back the tiles. The students also had to analyze all the possible moves that the computer could make so as to reduce the computer's chances of making a potentially game-winning move. Our results suggest that with more daily practice, the students could get even better at attempting activities that require them to use their logical reasoning and critical analysis skills under a time constraint. This practice would also directly result in an improvement in students' test scores, as students would have strengthened their skills after playing the games. This experiment could not be carried out for a longer duration, as the author was not able to obtain permission for a longer duration. A widespread implementation of such an activity is feasible, and many students can benefit from this activity. In the future, if permission is obtained, experiments will be carried out for a longer period of time in order to check long-term effects of the study. Further development of games will also be undertaken in order to improve the quality of the games so that the students find the games more challenging. The games will also include syllabus-related material to maximize the intersection of the studies with the games. This could help students learn concepts while doing something they love – gaming. Also, through such games direct results on test scores in school might also be observable.

**Methods**
The students were made to play the games in their first period of the school day. They were made to play these games for half an hour. Permission was obtained from the authorities to carry out the experiment on the school premises. The tests were carried out in the class and the students were made to play in the computer laboratory.

The games were built using Python with the help of Pygame. Pygame is a cross-platform set of Python modules designed for writing video games. It includes computer graphics and sound libraries to be used with the Python programming language. The games primarily tested the strategic skills of the students by forcing them to make decisions in real-time, with the aim of beating a computerized opponent. The students had to convert tiles on the board to the color that was assigned to them. The aim of the game was to choose a tile based on factors that would lead to conversion of more tiles and to convert strategic tiles, so as to prevent the loss of a large number of tiles when the computer plays its move. As the tiles could be converted back and forth, the students had to use their logical reasoning and critical analysis skills to carefully make a move that would increase their chances of preventing the computer from taking back the tiles. The student also had to analyze all the possible moves that the computer could make so as to reduce the computer's chances of making a potentially game-winning move. The students also had to plan their next move by analyzing the entire board and taking all the possible moves and their outcomes into consideration. The game can simply be explained as a hybrid between Othello and Go, two popular board games that can be played between two people. All the moves had to be completed under a time constraint. Failure to make a move within the time limit led to the move being passed to the computer.

The volunteers were seventh grade students from the Sardar Vallabhbhai Patel School. All the volunteers were from the same class. Along with the author, a teacher was present at all times to supervise the students and maintain discipline. A total of 45 students out of a class of 61 took part in the experiment. The students who did not complete either one of the tests were omitted from the study. Permission from the students and their guardians was obtained. No student was forcefully made to take part in this experiment.

The tests were of equal difficulty. The students were given half an hour to complete the exams. Maximum effort was taken in order to prevent students from cheating on the test. One test was given to the students prior to the experimentation period, while another test was given at the end of the week. The test papers were designed based on various competitive exams for the seventh grade. The questions were sourced from these exams, but not used exactly as they appeared in the original exam. All the data given in the questions were changed in order to make sure that students were given questions that they were not familiar with. The papers were solely created by the author. A tutor for competitive exams was consulted regarding the difficulty of the papers. It was ensured that the test was appropriate for seventh grade students. The questions primarily tested the logical reasoning and critical analysis skills of the students. Similar type of questions can also be found on tests conducted by Mensa, the Stanford-Binet test, etc. A question on the test might look something like the question given below:

The sum of the two 5-digit numbers XYZ10 and XYZ12 is 123422. What is the value of X + Y + Z?
$$(A) 10 \quad (B) 11 \quad (C) 12 \quad (D) 14$$

For the statistical calculations, the GraphPad Prism 6 trial version software for Mac was used. All calculations were either done manually or by the use of the aforementioned software.

**Acknowledgments**
The authors would like to thank the management at the Sardar Vallabhbhai Patel School for permission to carry out the experiments in the school premises and during the school hours. The authors would also like to thank all the volunteers who took part in this experiment. Last but not the least, the authors would like to thank their family for support throughout the course of this experiment.